# Smart Home Wireless Sensor Nodes

Addressing the Challenges using Smart Objects and Artificial Intelligence

P. Lynggaard
Center for Communication and Information Technologies
Aalborg University
Denmark
perlyn@es.aau.dk

*Abstract*—Smart homes are further development of intelligent buildings and home automation, where context awareness and autonomous behaviour are added. They are based on a combination of the Internet and emerging technologies like wireless sensor nodes. These wireless sensor nodes are challenging because they consume battery power, they use network bandwidth, and they produce wireless interferences. Currently, different methods exist for handling these challenges. These methods are, however, based on adjusting the transmitter frequency and using duty-cycling in combination with sleep mode approaches. This paper introduces an approach that considerably lowers the wireless sensor node power consumption and the amount of transmitted sensor events. It uses smart objects that include artificial intelligence to efficiently process the sensor event on location and thereby saves the costly wireless transportation of these events. In this paper it has been shown that this approach provides huge savings of power consumption and network load, which in turn reduces the interference level.

*Keywords*—*Smart homes, smart objects, artificial intelligence, embedded microprocessor models*.

## I. Introduction

Smart homes have their roots in the home automation area, which offers remote and timer control of systems and embedded devices such as light, heating, entertainment systems, and appliances to improve comfort, energy efficiency, and security. However, the elements of context sensing and autonomous behaviour are lacking, but these are offered by the smart homes. The context sensing is mainly provided by wireless sensor nodes, which are critical key components in smart homes. By providing processing power and Artificial Intelligence (AI) to these wireless sensor nodes they become Smart Objects (SOs).

The wireless sensor node challenges can be divided into two groups. Firstly, the huge amount of communicating wireless sensor nodes involved in smart homes produce interferences, which disturb other services. Conversely, the high interference level in modern homes reduces the sensor nodes communication ability. Secondly, the amount of sensor node information exchanged in smart home networks influences the sensor power consumption (battery lifetime), the interference level, and the use of low-power networks.

Based on these challenges a research question can be formulated as: How can "intelligent" SOs reduce interferences, bandwidth, and power consumption in smart home wireless sensor nodes?

This paper discusses and answers the research question by using an SO model, which has been implemented on a microcontroller based simulator. Using this simulator on a publicly available dataset from CASAS [1] it has been shown that the used spatially distributed SO concept provides huge savings in network traffic load and sensor node resources. Hence, the AI provided by the SOs means that instead of transmitting sensor events to a central smart home server, these are processed by the AI framework implemented in a locally placed SO. When the AI framework recognizes an event it emits an action and thereby transforms the huge amount of events into few actions. This transformation provides the savings.

The rest of this paper is organized as follows: Section II provides an overview of related works. Section III provides an overview of the SOs and smart home technologies. Section IV simulates and discusses the derived SO concept in the light of the discussed challenges. Section V concludes the paper by discussing the achieved results in relation to the challenges.

## II. Related Work

In today's homes many types of interferences arrive from the numerous deployed wireless devices. However, in smart homes even more wireless devices will be present in the form of sensors and actuators that make the smart home context aware.

Another major interference signal source in smart homes is a WiFi router that works in the industrial-scientific-medical band where most of the Smart Home Network (SHN) spectrum is allocated. Thus, WiFi routers will impose interference on the SOs. It has been shown that a common 802.15.4 based SHN will lower its capacity with 26 per cent in some WiFi based scenarios and that radiation from a microwave oven would cause packet losses of approximately eight percent in a radius of 1.5 meter [1], [2]. To overcome this Yao et al. [1] have proposed a concept where the sensor nodes adjust their transmitting frequency. A similar work by Hou et al. [2] confirms this problem. They recommended frequency adjustments based on an analytic model.

The consequences of these interferences are that the retransmission rate in the SHN will increase and the battery powered nodes are forced to increase their transmitter power and thereby their power consumption. Another important factor that increases the node transmitter power is shading, where the signal is lowered by disrupting furniture and walls [3]. Thus, the challenge is to reduce the interference level in smart homes for saving power in the battery driven sensors and SOs.

Power consumption in SOs is a problem, if they are powered by batteries [4], [5], [6]. Changing batteries on sensors and devices is a non-trivial task because they are often hidden in door frames or built into closets, i.e. they are hard to reach. This problem increases with the amount of available sensors that can be more than one hundred in a smart home [7]. Another issue is the physical size of the used batteries because they need to be small enough to fit into the low form-factor sensors and devices. An additional dilemma is that the sensor devices will become smaller in the future so smaller batteries with limited capacity are needed.

The challenge with battery power consumption in the SHNs has been discussed among researchers for years. Different approaches have been suggested such as an SO framework which is based on a battery powered master / slave approach [8] and a common SO gateway approach that saves battery power [9]. Jin et al. [10] have suggested a smart home network, which reduces sensor power consumption by lowering the bit-rate. Lutz et al. [11] have proposed a framework that saves battery power in sensor nodes by using a duty-cycling and a sleep mode approach.

### III. SMART OBJECTS AND SMART HOMES

To deal with the discussed challenges a distributed SO architecture has been derived. It is presented, simulated and discussed in a smart homes context.

#### A. Smart Objects

The basic principles behind the distributed smart home architecture are to combine the communication, the processing and most of the AI parts into small embedded devices, i.e. the SOs. These are then assigned simple atomic functionalities in the form of offering particular services to the user, i.e. actions. Limiting the functionalities saves resources and offers small form factors, why the SOs can be positioned close to their related sensors and they can be wired to these. The rationale is that when the user performs an action, it is bound to the sensors spatial context. Hence, when the user opens the fridge, takes a plate, sits down and eats breakfast all these sensor events are bound to the kitchen context.

From an AI point of view, an SO is able to learn and predict actions. An example could be a table lamp where the actions are light on, light off, and dim light, which requires three AI instances embedded in the table lamp SO. However, processing AI in SOs is challenging because it requires considerable processing, bandwidth, storage and power resources [12]. To overcome these challenges the AI algorithms and the sensor events have been optimized.

#### B. Smart Homes

Smart homes share common elements with the intelligent building and the home automation areas. Hence, the smart homes era builds on the progressing maturity of these areas and the Internet of Things evolution, adding artificial intelligence to the home automation field. Today's smart homes are mainly implemented on centralized servers [8]; however, it is expected that the next generation smart homes will be based on distributed architectures to overcome the technological challenges and to align with the IoT based future [12], [8]. Nevertheless, at present smart homes are still in their infancy, and they only exist in the form of laboratory experiments such as living labs [13].

The smart home infrastructure contains different network types [14]. One type is the well-known WiFi based Internet. Another type is the SHN that interconnects the sensors, the actuators and the cloud based user interface. This type uses low bit-rate networks like ZigBee, Bluetooth low energy, X10, INSTEON, and Z-Wave.

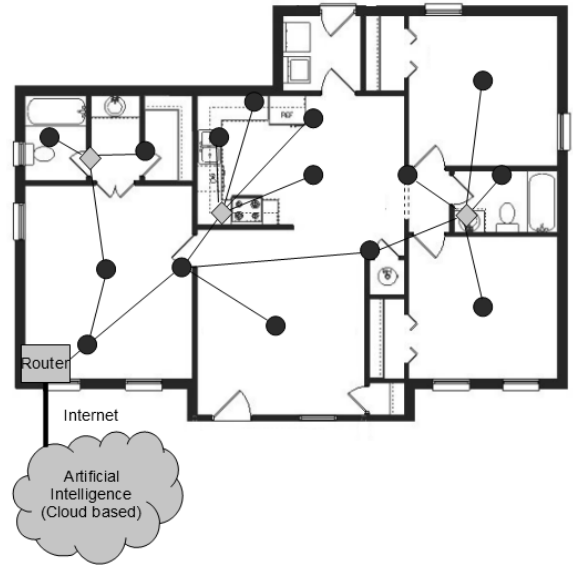

Fig. 1. Smart home equipped with sensor nodes. Light grey diamonds are smart objects.

A typical SHN that uses a transmit-all-event approach is illustrated in Fig. 1. The circles and diamonds are the sensor nodes, which are wirelessly connected (lines). When the sensors are activated they emit events that are routed across the home arriving at the Internet router. From the Internet router they are routed to the artificial intelligence system placed on a cloud server. This server receives all sensor events and it runs the AI algorithms, which predict and schedule services to the user. However, this work uses a distributed smart home architecture where the locally positioned SO nodes (diamonds) perform the AI processing. Processing the events locally reduces the number of transmissions significantly, which in turn reduces the interferences, the amount of exchanged sensor node information, and the overall battery power consumption.

### IV. SMART OBJECT SIMULATION

The smart object simulation is divided into two parts. First part presents the simulation setup, while the second part discusses the obtained results.

## A. Simulation Setup

An SO simulation model based on AI and simplified sensor information has been implemented and simulated on a state-of-the-art embedded microcontroller. This model has been used to analyze the discussed challenges by using a dataset from the WSU CASAS Milan smart home project [15]. This dataset contains sensor data collected in a smart home (Fig. 2) where a volunteer adult woman lived for three months. She had a dog and was visited by her children on several occasions. During the period, 26 sensors collected 433656 events and 2310 annotated actions. By routing five kitchen sensor events (circle in Fig. 2) into one SO it was possible to simulate and calculate the power and network load savings.

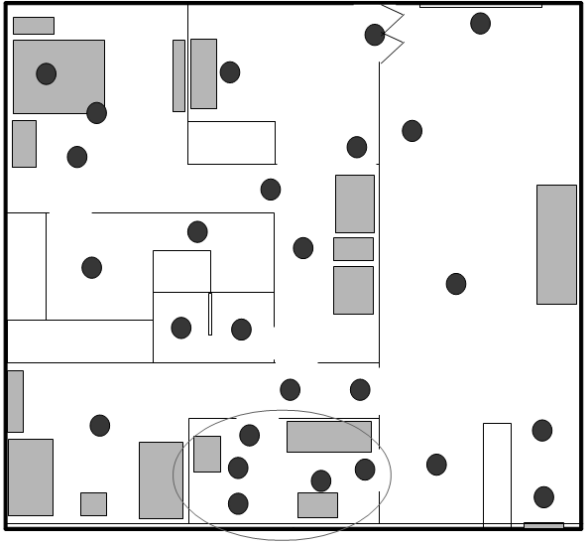

Fig. 2. WSU CASAS Milan smart home project [16]. Solid circles are smart home sensors. The large circle encloses the five kitchens sensors used in this work.

The used simulation model is illustrated in Fig. 3. It has been implemented in C-code using a state-of-the-art embedded device from Microchip. The implemented software inclusive test scripts have been executed in the MPLAB v1.51 simulator from Microchip. By using this environment it was possible to predict processing load and thereby indirectly calculate the consumed power.

The C-code was compiled and implemented on a low-power microcontroller PIC18F46J50 from Microchip. The processor clock speed was set to 1 MHz, where it consumed 1.2 mA when it was running and nothing else. Using this microcontroller family has some advantages: It is a cheap consumer product, it is supported by a comprehensive tool suite, and it offers very low power consumption, i.e. it can run from months to years on a small CR2032 battery depending on the processing load. In addition, it includes a variety of sleeping and processing modes together with built in hardware elements, such as timers, wakeup on port change interrupts, etc. All these features are useable in an SO component.

Implementation of the simulation model can be divided into two software parts: A First-In-First-Out (FIFO) buffer and an AI framework as illustrated in Fig. 3.

The FIFO (Fig. 3) uses a circular buffer approach which optimizes its efficiency. It contains a pointer "PTR" that always points to the oldest value. When a new event arrives, the interrupt routine in the microprocessor is activated. This interrupt routine handles a sequence of steps:

1. It fetches the new event and stores it in "Nval".

2. It fetches the oldest event from the circular buffer (pointed to by "PTR") and decrements the event counter (contained in "Acc") for that particular event.

3. It then adds the new event "Nval." to the event counter, i.e. the "Acc". Thus, the accumulator keeps track of how many instances for each sensor are present in the FIFO buffer.

4. It puts the new value into the circular buffer at the "PTR" position, i.e. it overrides the oldest value.

5. It then increments the pointer value to the next field, i.e. the (new) oldest value in the circular buffer.

6. It multiplies the accumulator with the weights and sums the products. Actually, it sums the weights that are multiplied by a binary one, i.e. no multiplication takes place. This provides efficiency.

7. It compares the summed weights with the threshold and emits an action if it is exceeded.

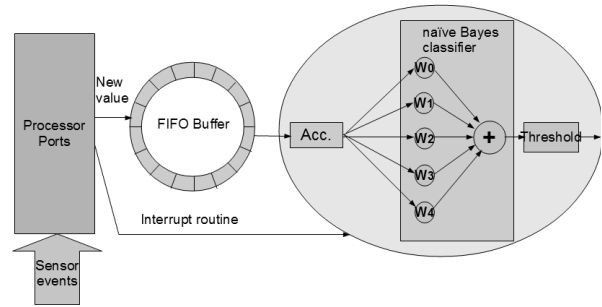

Fig. 3. Simulation model for the ai part of an SO. Left-hand is the buffer part and right-hand is the AI part.

A simplified overview of a SO and its included naïve Bayes based AI framework is provided in Fig. 4. The AI framework uses a naïve Bayes classifier, which is one of the most used probabilistic models in a smart home context [16], [17]. In addition, it provides a structure that allows adaptation to a simple sensor concept [18].

As discussed, supporting a distributed AI framework on an SO embedded platform requires heavy resources such as network bandwidth, processing power, and storage capacity [12], [19]. To overcome these challenges the employed naïve Bayes classifier algorithms have been simplified and optimized with respect to: It supports simplified binary sensor events, it uses the off part of the sensor events only, and it performs simple integer arithmetic.

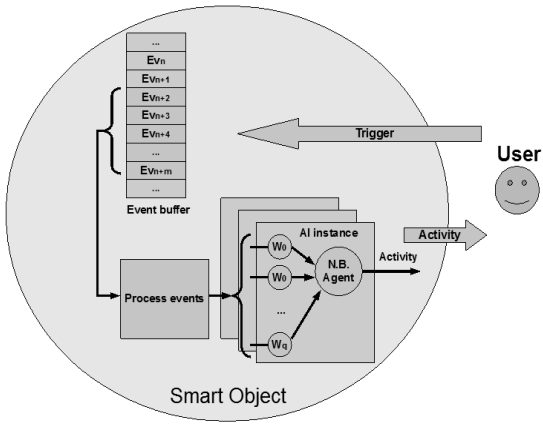

Fig. 4. Artificial Intelligence framework embedded in the smart objects.

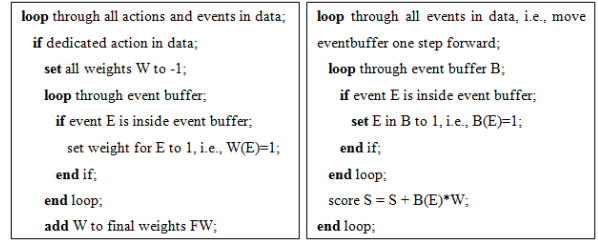

Fig. 5. Pseudo code for the essential algorithms in the naïve Bayes simulator. Left-hand code is the learning algorithm and right-hand code is the prediction algorithm.

The AI framework (Fig. 4) runs in a forever manner, monitoring its input for available sensor events. From an implementation point of view, this means that it powers down and waits for wakeup interrupts emitted when sensor data arrive. Thus, the prediction process is performed every time a sensor event arrives. It puts the sensor events into the event buffer. From this buffer, a group of sensor events is fed to a process event transformer. It combines the input sensor events and runs a balance system that prevents the AI weights from overflowing (i.e. scaling). Output from the transformer is multiplied by the weights $W_0$ and its approximated logarithmic values are summed into one value. This value is compared with a dynamic threshold, limited, and a binary decision is made as to whether or not to emit the detected user action.

Similar to the prediction process, the process of user activity learning occurs every time the action to be learned arrives. This action is marked in the sensor event buffer and all sensor events present in the buffer are processed by the event transformer. Output from this transformer is used to update the weights.

The AI instance (Fig. 4) represents the used naïve Bayes classifier framework. Its implemented behaviour is presented in form of a descriptive pseudo-code as illustrated in Fig. 5. The left-hand part provides pseudo code that covers the learning algorithm functionality. It loops through all sensor events when a defined user action arrives and it updates the weights accordingly. At the right-handed part the pseudo-code covering the prediction process is presented. Every time that a new sensor event arrives, its buffer representation is set to a binary one. By multiplying this one with the respective weight and finally adding all of these provide a score. This score is then threshold; if it is above, the dedicated action is claimed to be evident.

However, it is noted that this pseudo code only illustrates the main functionality and does not include all the necessary handling, such as scaling and presenting the data.

The AI part embedded in SOs needs user triggered sensor events from its context. Thus, smart homes sensors are the primary source for information of user activity. Most sensors in smart homes are binary of nature, i.e., they only behave in an on-off manner. They offer simplicity and low network loads because only binary values need to be transferred. It is assumed that these sensors are placed in the smart home at optimal positions and that they are able to transmit simple events with only a small delay (less than a fraction of a second). These simple events contain a limited amount of information (identity) and they are expected to occupy only a few bytes of payload in the transmission context. It is also assumed that each sensor samples its context every second and that this sampling period is fast enough to avoid aliasing (i.e. it obeys the Nyquist-Shannon sampling theorem). Therefore, from a transmission point of view, every second, only a limited amount of information is sent from each sensor when it is activated; for the wireless sensors, this keeps the network load low. Such a simplification comes at a price because real-time continuous event is quantized into an impulse event, which is delayed. Another simplification used in this work is using the on-event only, i.e. the off-event is not transmitted. This choice comes at a price because the event duration information is also lost. However, it has been shown that such a concept does not degrade the recognisability significantly [20].

B. Simulation analysis

To be able to compare the savings in the transmit-all-events and the SO based approaches the energy consumptions for the two systems need to be estimated. Calculating this energy consumption requires that the system activity timeslots and the power consumption in these timeslots are known. So these parameters will be derived in the following.

By using the SO simulation model (Fig 3) the activity timeslots for the SO based approach can be found. Its implementation has been optimized for time so instead of traversing the full event buffer, which takes 38.5 mS for 1800 values (simulated values), it uses the accumulator balancing approach illustrated in Fig. 3. This approach performs the same task in 0.630 mS and thereby provides approximately 98 % time saving.

Whenever an event arrives (Fig. 3) it generates an interrupt, which runs the FIFO tasks in 0.630 mS. Added to this is the AI task (Fig. 4) that adds the five weights in 0.346 mS. Therefore, in total, the processor load for performing one agent instance prediction calculation based on five sensors is 0.976 mS. Assuming the sensor sampling frequency is 1 Hz and that they fire asynchronously the total processor load is below 0.5 %. It is noted that this is the maximum load because all sensors are changing all the time, but this is not often the case in a real-world scenario. As an example this means that 100 AI (i.e. SO) instances each connected to five sensors will use approximately 10 % of the processing resources. Because the AI processing is the main activity in an SO, it also means that the processor will be able to save power by sleeping for approximately 90 % of the time. As noted, these calculations are performed for the SO prediction process because it is more resource consuming than the learning process.

As discussed, a simplified NB algorithm has been used in the embedded SO model implementation for providing load saving. Thus, a non-optimized NB algorithm uses a processing calculation time of 17.53 mS that should be compared to the earlier discussed results of 0.976 mS for the optimized algorithm.

The savings by using the optimized buffer and the simplified NB algorithms is illustrated in Fig. 6.

As discussed, smart homes use wireless sensors, which are mostly based on low-power ZigBee transceiver technology [21]. Thus, for the following energy calculation, it is assumed that the sensors and SO nodes use the popular CC2420 ZigBee communication unit from Texas instruments [22], [23].

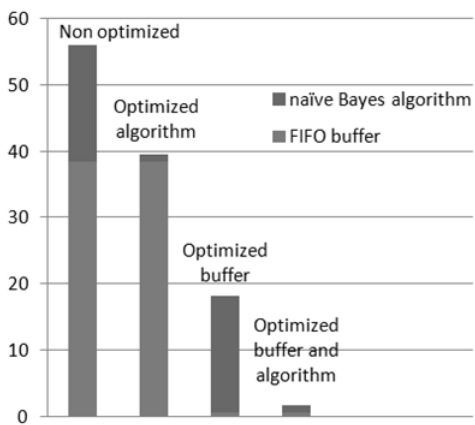

Fig. 6. Processing time (in ms) by optimizing the FIFO buffer and the naïve Bayes algorithm (y-axis is in milliseconds). Four results are presented: Non-optimized, optimized algorithm, optimized buffer, and optimized buffer and algorithm.

In a traditionally centralized setup (transmit-all-events) the five sensors in the CASAS smart home kitchen, [13] must transmit events every time they are triggered. Therefore, finding the cost of transmitting one sensor event and multiplying this by the average number of sensor events from these five sensors give the total cost per chosen unit time. Thus, an estimate of the average number of sensor events can be calculated by using the CASAS dataset and looking up the five sensor events over a time unit. It has been found that the average number of sensor events from these five sensors on a daily basis, is 1795 events.

Based on the calculated number of events the energy consumption can be found. It is assumed that only a three byte sensor identity is sent in each event and that the ZigBee transmitter uses 32.5 mA for the carrier sense multiple access sequence of length 2.9 mS, 13mA in 13 mS for activating the microcontroller, and 30.5mA in 1 mS for actually transmitting [4]. Adding up these contributions and multiply it with the average number of events and a 2-volt supply voltage (minimum operation voltage for the ZigBee transceiver) gives an energy consumption of 1.055 Joule per day. Assuming the sensor nodes are powered by a standard CR2032 battery that offers approximately 0.675 Joule means that these five sensors in total will use approximately 1.5 batteries per day.

However, using an SO to handle the five sensors as discussed earlier results in a power consumption of 1.2 mA for the processor running in 0.976 mS, which yields 2.34 uJ per event. Therefore, processing 1795 events costs 4.2 mJ. Assuming that the SO sends all the average detected kitchen actions (approximately 7 per day in the dataset) to a user interface system means that the cumulative transmitter consumption is 4.1 mJ. The final SO consumption is 8.3 mJ per day, which means that a CR2032 battery will last for 81 days. Comparing the SO energy consumption to the discussed "tx all event" scenario provides a saving of approximately 99%.

As discussed, for the simulations the following settings have been used: Processor clock speed 1 MHz, processor power consumption 1.2 mA when it was running and nothing when it was idling and each event consisted of a three byte sensor identity. Additionally, it was assumed that the network nodes used ZigBee devices. Regarding the network load the number of ZigBee frames or Internet packets is reduced from 1795 to 7 per day. These results are illustrated in Fig. 7 and Fig. 8.

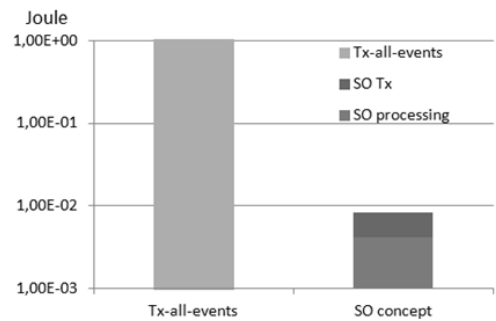

Fig. 7. A traditional sensor framework (Transmit-all-events) is compared with the presented SO concept. Five kitchen sensors in a CASAS smart home [20] have been used.

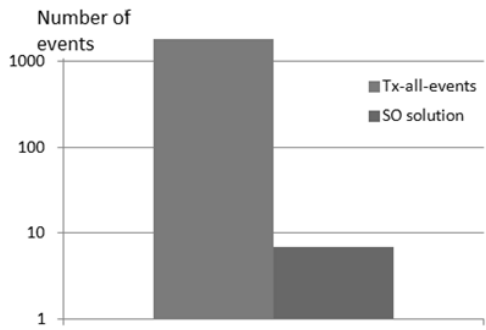

Fig. 8. Number of transmitted events in a traditional sensor framework (Transmit-all-events) and the presented SO concept. Light grey is Tx-all-events and dark grey is SO based solution.

Regarding the interferences it is noted that the presented result uses an average number of 1795 sensor events per day. However in real life, these sensor events are not distributed equally over a day because most people sleep at night and thereby concentrate the activities triggering the sensors to the daily hours as shown in Fig. 9. Thus, as shown, the savings in interference patterns will actually be larger than the discussed estimation during the time daily hours, i.e., from 7 to 21.

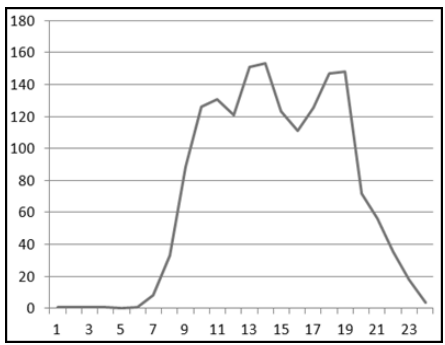

Fig. 9. Event-distribution for the simulated CASAS SMART HOME [9] (five kitchen sensors). the x-axis is daily hours and the y-axis is number of events per day.

## V. CONCLUSION

From these simulated results some conclusions can be drawn. Firstly, the interference level in smart homes is reduced considerably. Thus, reducing the 1795 routed sensor events to only 7 provides a huge reduction of interference level. Secondly, the amount of sensor node information exchanged is reduced likewise, i.e. the used bandwidth is lowered. Thirdly, the overall sensor node power consumption is reduced by 99 per cent. Finally, the concept of transmitting sensor events every second and only transmit sensor turn-on events provide additional savings.

However, the cost of these savings is the use of complex SO nodes with embedded AI. This AI needs training to achieve a detection probability comparable to similar systems.